# Anatomically and mechanically conforming patient-specific spinal fusion cages designed by full-scale topology optimization

Thijs Smit[1,*], Niels Aage[2], Daniel Haschtmann[3], Stephen J. Ferguson[1], Benedikt Helgason[1]

1. Institute for Biomechanics, ETH Zürich, Zürich, Switzerland.
2. Solid Mechanics, Technical University of Denmark, Denmark
3. Department of Spine Surgery and Neurosurgery, Schulthess Klinik, Zürich, Switzerland

*thsmit@ethz.ch

## Abstract

Cage subsidence after instrumented lumbar spinal fusion surgery remains a significant cause of treatment failure, specifically for posterior or transforaminal lumbar interbody fusion. Recent advancements in computational techniques and additive manufacturing, have enabled the development of patient-specific implants and implant optimization to specific functional targets. This study aimed to introduce a novel full-scale topology optimization formulation that takes the structural response of the adjacent bone structures into account in the optimization process. The formulation includes maximum and minimum principal strain constraints that lower strain concentrations in the adjacent vertebrae. This optimization approach resulted in anatomically and mechanically conforming spinal fusion cages. Subsidence risk was quantified in a commercial finite element solver for off-the-shelf, anatomically conforming and the optimized cages, in two representative patients. We demonstrated that the anatomically and mechanically conforming cages reduced subsidence risk by 91% compared to an off-the-shelf implant with the same footprint for a patient with normal bone quality and 54% for a patient with osteopenia. Prototypes of the optimized cage were additively manufactured and mechanically tested to evaluate the manufacturability and integrity of the design and to validate the finite element model.

**Highlights**

- A novel full-scale topology optimization formulation tailored for patient-specific spinal fusion cages, taking the structural response of the adjacent vertebrae into account in the optimization process.

- The optimized patient-specific spinal fusion cages result in anatomically and mechanically conforming devices.

- Anatomically and mechanically conforming devises reduce subsidence risk compared to other cages, by lowering strain concentrations in the adjacent bone structures.

- The reduction in subsidence risk is 91% compared to an off-the-shelf implant with the same footprint for a patient with normal bone quality and 54% for a patient with osteopenia.

**Graphical abstract**

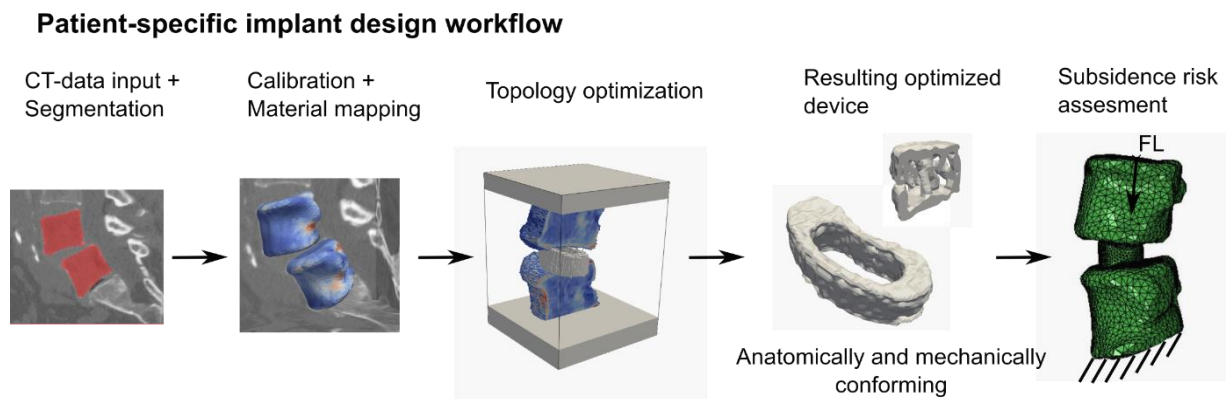

**Patient-specific implant design workflow**

CT-data input + Segmentation → Calibration + Material mapping → Topology optimization → Resulting optimized device / Anatomically and mechanically conforming → Subsidence risk assesment

# 1 Introduction

Instrumented spinal fusion surgery using intersomatic cages is a common treatment for degenerative spondylolisthesis or degenerative disc disease [1]. Failure of the treatment can lead to revision surgeries. A major contributor to failure of this procedure is cage subsidence, where the cage penetrates the vertebrae endplates due to over-loading of the endplates and underlaying bone structures [2]–[4]. This clinical problem is prominent for the Transforaminal Lumbar Interbody Fusion (TLIF) surgical approach due to the smaller cage sizes which is limited by the narrow access channel to the disc space proximate to the nervous structures [5].

Recent advancements in computational techniques and Additive Manufacturing (AM), make patient-specific spinal fusion cages accessible. Mobbs *et al.* treated several patients with Anatomically Conforming Devices (ACDs) where the cage's geometry was adapted to match the patient's vertebral endplate shape, using Computed Tomography (CT) scans as input [6]. These ACDs were suggested to reduce subsidence risk compared to Off-The-Shelf (OTS) cages [7]. Recent biomechanical studies confirmed this observation by evaluating the subsidence risk of ACDs in comparison to OTS cages in experimental and numerical studies [8]–[11]. More specifically, Fernandes *et al.* showed that the increased contact area between the cage and the endplate, by using an ACD (1.06 MPa) compared to an OTS cage (1.84 MPa), could reduce contact stresses at the endplates.

Topology optimization (TO) is a numerical method for iteratively optimizing the material lay-out within a design domain to best achieve specific functional goals [12]. TO has been used to optimize OTS spinal fusion cages for the lumbar and the cervical spine, respectively [13]–[17]. One of the motivations for using TO is to mitigate the risk of subsidence. Several authors optimized the implant's mechanical and morphological properties resulting in porous implant structures with the aim to facilitate bone in-growth [18], [19]. Bone in-growth is important for secondary stability of the implant. Primary stability is provided by the implant before bone ingrowth is achieved. During the post-operative period before secondary stability is achieved, avoiding subsidence is critical.

Previous studies employing TO for optimizing OTS cages do not incorporate patient-specific information. Current ACDs are designed to match the patient's anatomy but ignore the load bearing capacity of the adjacent bone structures. Furthermore, treating patients with low bone quality, using OTS cages, leads to relatively high revision rates [20]. The structural information of the adjacent vertebrae can be included in the implant TO process, to enable patient-specific structural optimization, resulting in anatomically and mechanically conforming devices with the potential to further reduce subsidence risk.

Thus, the aim of this study was to optimize patient-specific spinal fusion cages that reduce subsidence risk by taking the structural response of the adjacent vertebrae into account. We generated patient-specific spinal fusion cages with a novel full-scale TO formulation, denoted by Topologically Optimized Devices (TODs), and compared the subsidence risk of the TOD to a currently used OTS cage and an ACD. We hypothesized that anatomically and mechanically conforming devices (AMCDs) reduce subsidence risk compared to ACDs and OTS cages, by lowering strain concentrations in the adjacent bone structures.

## 2 Methods

### 2.1 Study subjects

Anonymized pre-operative CT scans with clinical resolution (voxel sizes of about: 0.30 mm × 0.30 mm × 0.50 mm) of two human Functional Spinal Units (FSU), from two female patients, were provided by the Schulthess Klinik, Zürich (Table 1). These patients were diagnosed with degenerative spondylolisthesis and treated with a TLIF procedure. The patients had no history of spinal diseases, spinal trauma, or spinal deformity. Based on the available T-score (from DXA scan) and the calculated mean integral volumetric bone mineral density (integral vBMD), patient 1 was classified as osteopenic and patient 2 was classified as having normal bone quality. The integral vBMD war calculated using the apparent density $\rho_{app}$, including the endplates and cortical bone [21].

*Table 1, patient selection and information, NA: not available (all the subjects gave informed consent).*

| FSU ID | Age | Gender | BMI | Weight [kg] | Level | vBMD [g/cm$^3$] | T-Score | Classification |
|--------|-----|--------|------|-------------|-------|-----------------|---------|----------------|
| 1 | 66 | Female | 18.3 | 51 | L4 | 0.16 | -2.3 | Osteopenic |
| | | | | | L5 | 0.15 | NA | |
| 2 | 58 | Female | 25.6 | 75 | L4 | 0.40 | NA | Normal |
| | | | | | L5 | 0.45 | NA | |

## 2.2   Model building

As the CT scans were not calibrated during the image acquisition, we employed a phantom-less calibration method that was previously published and results in scan-specific linear calibration equations (Figure 1.a) [22].

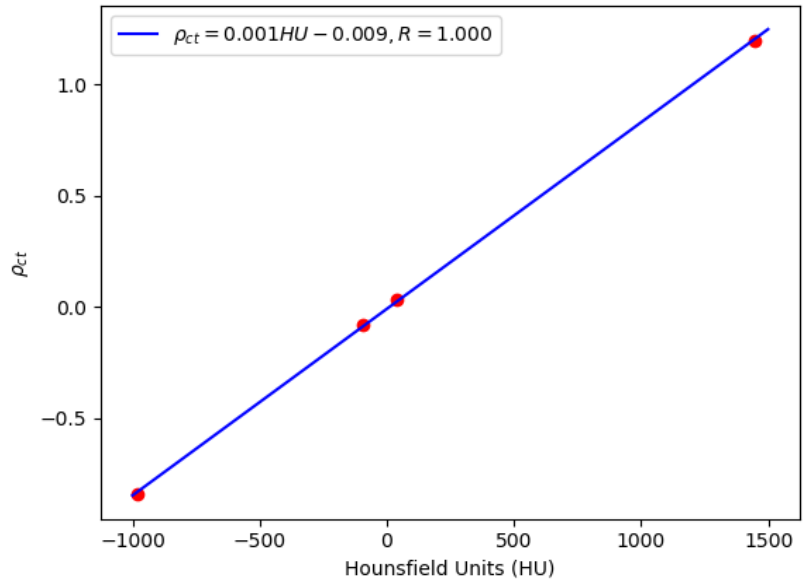

**(a)**

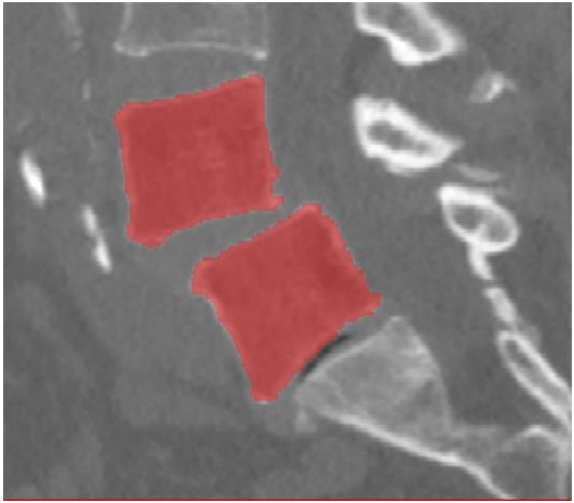

**(b)**

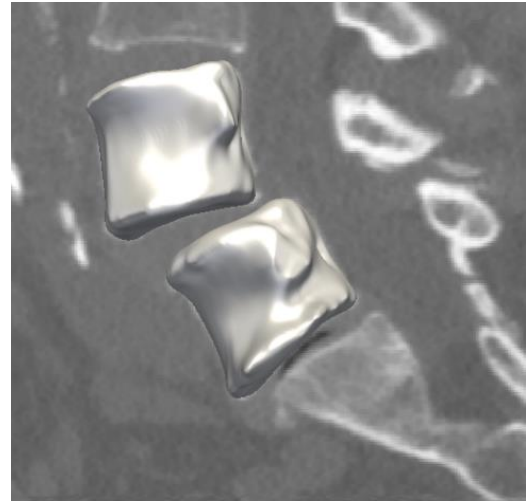

**(c)**

*Figure 1, a) scan-specific linear calibration equation to map the Hounsfield Units to $\rho_{ct}$ b) segmentation slice in MITK-GEM c) surface representation of the segmented vertebrae (all data from patient 2).*

Segmentation of the vertebrae was performed manually using the open-source software MITK-GEM [23] (Figure 1.b and Figure 1.c). The material mapping function from *Ouyang et al.* [24] to obtain the Young's moduli $E(\rho_{app})$ was chosen because it is a well validated and the best performing mapping function for vertebral bodies according to previous studies [25], [26]. An overview of the variables and their relationship is provided in Table 2. The Poisons ratio was set to 0.3 and the minimum value for the Young's modulus was 25 MPa.

*Table 2, variables related to calibration and material mapping based on CT scans.*

| Parameter | Function | Unit | Source |
|---|---|---|---|
| $\rho_{ct}$ | Linear relationship from calibration | g/cm$^3$ | [22] |
| $\rho_{ash}$ | $0.0789 + 0.877 * \rho_{ct}$ | g/cm$^3$ | [27] |
| $\rho_{app}$ | $\rho_{ash}/0.6$ | g/cm$^3$ | [25] |
| $E$ | $2583 * \rho_{app}^{1.88}$ | MPa | [25] |

A commonly used OTS commercial banana cage for a TLIF procedure with a height of 9 mm was reengineered into a CAD model (Figure 2.a). This CAD model was used as the basis for the ACD and the TOD design domain by extruding the top and bottom surfaces to match the patient's endplates (Figure 2.b).

Intervertebral disc material and cartilaginous endplates were assumed to be removed during the surgical procedure. The cages were placed anteriorly and across the midline in an optimal position, restoring spinal alignment as much as possible (Figure 2.c and Figure 2.d). The implant placement was checked by a spine surgeon according to standard surgical fusion techniques for the TLIF approach. The ACD model was used as a design domain for the TO process (section 2.3). This ensured that the contact areas between the cages and endplates were equal for all models to allow a fair comparison. The .stl files of the ACD and the OTS cage were saved for use in the subsidence risk assessment (section 2.4).

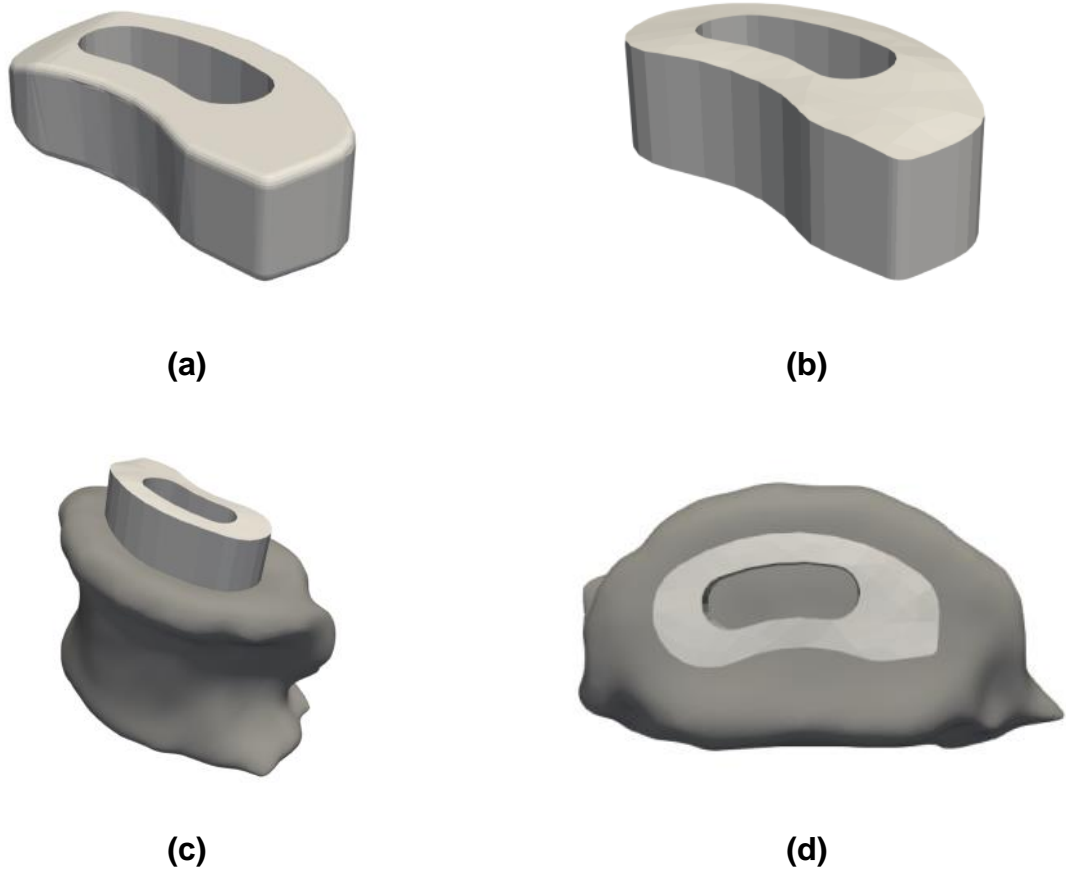

*Figure 2, a) CAD model of 9 mm in height OTS cage, b) CAD model of the ACD, c) ACD positioning relative to the inferior vertebra L5, d) ACD positioning relative to the inferior vertebra L5 (all data from patient 2).*

## 2.3   Topology optimization

Having processed the patient-specific data, the following sections present how TO was applied to the problem. The optimization domain $\Omega$ was composed of an implant domain ($\Omega_{implant}$), a bone domain ($\Omega_{bone}$) and a rigid domain ($\Omega_{rigid}$), with $\Omega = \Omega_{implant} \cup \Omega_{bone} \cup \Omega_{rigid}$ (Figure 3). $\Omega_{bone}$ contains the bone structures and patient-specific material mapping which are obtained from the patient's CT scans. $\Omega_{implant}$ is the design domain that is used in the TO process.

The full-scale topology optimization formulation was implemented using a large-scale TO framework which can design structures with high-resolution [19], [28], [29]. The formulation follows the density approach with a linear elastic Finite Element (FE) model, assuming small strains. The Solid Isotropic Material and Penalization method (SIMP) is used and as optimizer the Method of Moving Asymptotes (MMA) is integrated [30].

Domain Ω was discretized using a structured grid of 1.728 million hexahedron elements with a mesh size of 0.45 mm. A resampling was performed to adjust the CT scan resolution to Ω using SimpleITK version 2.1.1.2 [31]. The bone-implant interface was assumed to be fully bonded. A PDE filter was used to avoid numerical artifacts [32].

Titanium was selected as cage material as it is widely used in orthopaedic implants [33]. Furthermore, metal powder bed fusion processes have sufficiently high-resolution to additively manufacture the optimized implants.

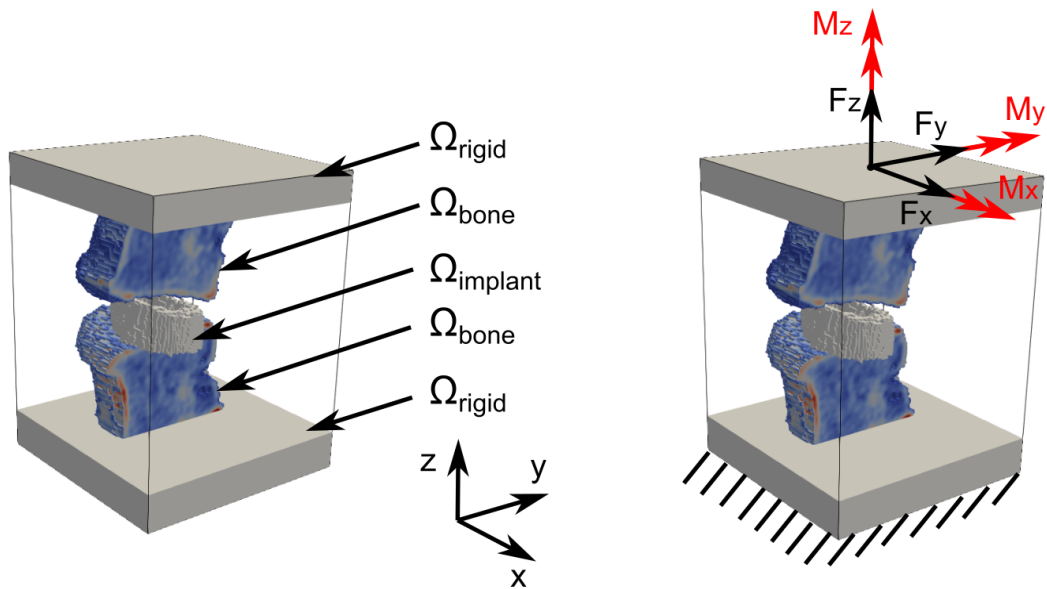

*Figure 3. (Left) computational domain Ω with Ω = $\Omega_{implant} \cup \Omega_{bone} \cup \Omega_{rigid}$. (Right) three forces, in black, and three moments, in red, are applied on top of Ω in the centre. Support constraints are applied to the bottom of Ω (data from patient 2).*

A two-stage design process was employed. In the first stage the cage was optimized for daily-living loading conditions. These daily-living loads, because they are relatively small, cause strains in the adjacent bone structures and the implant in the linear regime. For the proposed optimization problem, the use of a linear elastic FE model was a reasonable and sufficient assumption, as the adjacent bone structures and implant do not undergo plastic deformation under daily-living loading in the optimization process. In the second stage, a verification was performed in a commercial finite element analysis software, where the cages were loaded with hyper-physiological loading, to evaluate and compare the implant's subsidence risks. In this stage a non-linear FE model and material model was used to quantify the post-yield effects and subsidence risk (section 2.4).

Six load-cases (axial compression, lateral shear, posterior-anterior shear, flexion, lateral bending, and axial rotation) were used for the daily-living loads, based on the study of *Rohlmann et al.* and the Orthoload dataset [34], [35] (Figure 3). For each load-case, a recorded daily-living loading out of the Orthoload dataset was used, from a patient matched based on body weight (Table 3) [36]. Loads were applied on nodes at the top of $\Omega$ and nodes on the bottom of $\Omega$ were fixed in all directions (Figure 3). The difference between the point of measurement of the loads in *Rohlmann et al.* and the load application point on $\Omega$ was accounted for by adjusting the applied loads to match the experimental results of Rohlmann *et al.*

*Table 3, spinal loads during daily living applied in the TO process, derived from the study of Rohlmann et al. and the Orthoload dataset* [34], [35].

| Description | Component | Patient 1 | Patient 2 | Unit |
|---|---|---|---|---|
| Axial compression | $F_z$ | -177 | -390 | N |
| Posterior/Anterior shear | $F_y$ | 27 | 60 | N |
| Lateral shear | $F_x$ | 5 | 12 | N |
| Flexion | $M_x$ | 0.4 | 0.9 | Nm |
| Lateral bending | $M_y$ | 0.5 | 1.2 | Nm |
| Axial rotation | $M_z$ | 0.7 | 1.5 | Nm |

The most important function of the spinal fusion cage is to provide primary stability to the treated segment, allowing the two vertebrae to fuse together [37]. From an optimization perspective, this means that we aim to maximize stiffness of the bone-implant system for all load-cases (optimization objective) but also avoid over-loading of the adjacent bone structures. To this end, the maximum and minimum principal strains of the bone structures in the adjacent vertebrae ($\Omega_{bone}$), were constrained (optimization constraints). Apart from strain constraints, we employed a local volume constraint on $\Omega_{implant}$ to create a porous implant structure and control the porosity and pore-size distribution to be favourable for bone in-growth. To promote manufacturability of the designed cages by a metal powder bed fusion process, a minimum strut diameter control was used via a robust formulation. Additionally, a connectivity constraint was added to the formulation and applied to $\Omega_{implant}$, to create an interconnected porosity which facilitates bone in-growth and the removal of unsolidified powder in the additive manufacturing process.

To ensure the optimized designs are robust wrt. manufacturing as well as strain levels in $\Omega_{bone}$, we employed a so-called robust topology optimization formulation. This amounts to the application of a modified robust optimization formulation, based on the three-field density projection; the eroded $x^e$, the intermediate $x^i$ and the dilated design $x^d$ [38]. The compliance minimization problem with maximum and minimum principal strain constraints, local volume constraint and connectivity constraint in discrete form is written as:

$$\min_{x} \sum_{k=1}^{lc} f(x^e)_k \quad in \quad \Omega$$

          1, objective function

$$subject\ to$$

$$\sum_{k=1}^{lc} g(x^d)_k \leq 0.0 \quad in \quad \Omega_{bone}$$

          2, maximum principal strain constraint

$$\sum_{k=1}^{lc} q(x^d)_k \leq 0.0 \quad in \quad \Omega_{bone}$$

          3, minimum principal strain constraint

$$v(x^d) \leq 0.0 \quad in \quad \Omega_{implant}$$

          4, local volume constraint

$$c(x^d) \leq 0.0 \quad in \quad \Omega_{implant}$$

          5, connectivity constraint

where $lc$ represents the six load-cases with $k = 1 \dots 6$, $f$ the objective function, $g$ and $q$ the constraints on the maximum and minimum principal strain response in the $\Omega_{bone}$, respectively, $v$ the local volume constraint and $c$ the connectivity constraint on $\Omega_{implant}$.

The objective was to maximize the stiffness (minimize compliance) of the bone-implant system in $\Omega$ for all load-cases. The objective function is evaluated using the eroded design $x^e$ because this realization is producing the worst-case objective values. The objective function is defined as the strain energy corresponding to load-case $k$:

$$f(x^e)_k = (u^T K u)_k$$

where $u$ is the displacement field corresponding to load-case $k$, and $K$ the stiffness matrix. The maximum and minimum principal strains are calculated using:

$$pe\_max = \frac{2}{\sqrt{3}}\sqrt{I_2} \sin\left(\theta + \frac{2\pi}{3}\right) + \frac{I_1}{3} \qquad 7$$

$$pe\_min = \frac{2}{\sqrt{3}}\sqrt{I_2} \sin\left(\theta - \frac{2\pi}{3}\right) + \frac{I_1}{3} \qquad 8$$

where

$$\theta = \frac{1}{3}\sin^{-1}\left(-\frac{3\sqrt{3}}{2}\frac{I_3}{I_2^{3/2}}\right), -\frac{\pi}{6} \le 0 \le \frac{\pi}{6} \qquad 9$$

The strain invariants $I_1$, $I_2$ and $I_3$ are calculated using:

$$I_1 = \mathbf{M}\,\boldsymbol{\varepsilon} \qquad 10$$

$$I_2 = \frac{1}{3}\boldsymbol{\varepsilon}^T \mathbf{V}\boldsymbol{\varepsilon} \qquad 11$$

$$I_3 = s_{11}\,s_{22}\,s_{33} + 2s_{23}s_{13}\,s_{12} - (s_{11}s_{23}^2 + s_{22}s_{13}^2 + s_{33}s_{12}^2) \qquad 12$$

where $s_{ij}\ i,j = 1,..,3$ are the components of the deviatoric strain tensor and $\boldsymbol{\varepsilon}$ the element strain tensor $[\varepsilon_{11}\ \varepsilon_{22}\ \varepsilon_{33}\ \varepsilon_{23}\ \varepsilon_{13}\ \varepsilon_{12}]^T$, which is obtained by averaging the contributions from the 8 integration points per element. The deviatoric strain tensor $\boldsymbol{s}$ is computed by:

$$\mathbf{s} = \boldsymbol{\varepsilon} - \boldsymbol{\varepsilon}_{hyd} \qquad 13$$

$$\boldsymbol{\varepsilon}_{hyd} = \frac{\varepsilon_{11} + \varepsilon_{22} + \varepsilon_{33}}{3} \qquad 14$$

Furthermore, $\mathbf{M} = [\,1\ 1\ 1\ 0\ 0\ 0\,]$ and

$$\mathbf{V} = \begin{bmatrix} 1 & -1/2 & -1/2 & 0 & 0 & 0 \\ -1/2 & 1 & -1/2 & 0 & 0 & 0 \\ -1/2 & -1/2 & 1 & 0 & 0 & 0 \\ 0 & 0 & 0 & 3 & 0 & 0 \\ 0 & 0 & 0 & 0 & 3 & 0 \\ 0 & 0 & 0 & 0 & 0 & 3 \end{bmatrix} \qquad 15$$

The constraint values, before aggregation are introduced using:

$$\Lambda_{max} = \frac{pe\_max}{pe\_max_{lim}} - 1.0 \qquad 16$$

and

$$\Lambda_{min} = \frac{pe\_min}{pe\_min_{lim}} - 1.0 \qquad 17$$

Here, $\Lambda_{max}$ and $\Lambda_{min}$ are aggregated using a p-mean aggregation function to obtain $g$ and $q$:

$$g = \frac{1}{n} \left( \sum_1^n \Lambda_{max}^{\ p} \right)^{1/p} \qquad 18$$

$$q = \frac{1}{n} \left( \sum_1^n \Lambda_{min}^{\ p} \right)^{1/p} \qquad 19$$

The constraints, $g$ and $q$, are calculated using the dilated design $x^{\mathrm{d}}$ as this realisation is producing the worst-case strain states in $\Omega_{\text{bone}}$. A maximum principal strain limit $pe\_max_{lim}$ of 0.73% and a minimum principal strain limit $pe\_min_{lim}$ of -1.04% [39] were used to limit the stain states in $\Omega_{\text{bone}}$.

After producing results using the formulation, we realized that the maximum principal strain constraint for the compression load-case was dominant. To save computational cost we deactivated inactive strain constraints during the TO process but checked all strain constraints every 25 iterations. A p-mean penalty value of 22 was used. For more details on a similar formulation see [40].

To ensure mesh-independent solutions to the posed optimization problem, a length scale must be introduced. This is obtained by filtering schemes. A PDE filter [32] converts the mathematical design variables $x$ to the filtered design variables $\tilde{x}$. A projection is used to ensure a solid/void design. The projection results in the physical design variable $\hat{x}$ [41] with,

$$\hat{x} = \frac{\tan^{-1}(\beta\mu) + \tan^{-1}(\beta(\tilde{x} - \mu))}{\tan^{-1}(\beta\mu) + \tan^{-1}(\beta(1 - \mu))} \qquad 20$$

with $\beta$ the projection parameter and $\mu$ the threshold parameter corresponding to realization $\mu^{d}, \mu^{i}$ and $\mu^{e}$. $\beta = 1$ was used at the start of the optimization and increased to a maximum of $\beta = 64$.

The use of the robust formulation resulted in a controllable minimum strut diameter which improves manufacturability and the integrity of the implant structure. Having at least 4 elements through the cross section of a single strut was targeted (minimum length-scale $r_{min}$). Therefore, a filter radius $r_{fil}$ is set to 0.40. $\mu^{e} = 0.65$ was used to generate the eroded design and $\mu^{d} = 0.35$ to generate the dilated design. For the intermediate design $x^{i}$ we used $\mu^{i} = 0.5$. This lead to the following relation between the minimum length-scale $r_{min}$ and the filter radius $r_{fil}$ for the solid and void phase [41]:

$$r_{min} = 2 * r_{fil} * 1.15 \qquad 21$$

The local volume constraint $v(x^{d})$ is followed the implementation published by Wu *et al.* [42]. The local volume constraint was applied on the dilated design $x^{d}$, with the local volume fraction to be 0.35, the filter radius of 3.5 and p-mean penalty of 16.

A connectivity constrain, $c(x^{d})$, based on the effective stiffness constraints from Swartz *et al.* [43], was added to the formulation. The mathematical design variable field is inverted with $x_{inv} = 1 - x$. Thereafter, an inverted filtered and dilated variable field is created, *i.e.*, $\tilde{x}_{inv}$ and $x^{d}_{inv}$, respectively. A compression load and a minimum stiffness (or maximum compliance) constraint is applied in the form $U(x^{d}_{inv}) \leq \gamma$. With $\gamma$, a user defined parameter which need to be tuned. $U(x^{d}_{inv})$ is the strain energy of the inverted dilated density field in $\Omega_{implant}$. Decreasing $\gamma$ would increase the size of the void regions in the final design and increase the likelihood of interconnected void spaces to facilitate the removal of unsolidified powder during the AM process. Finally, the Poisson´s ratio for titanium was set to 0.3, the E-modulus to 110 GPa [44] and the SIMP penalty and homogeneous initial condition were fixed at 3.0 and 0.35, respectively.

The optimization was run for 24 hours on a computational cluster using 1000 cores in parallel. The final design should at least exhibit a discreteness measure that is lower than 3%. This value is considered to be an acceptable convergence to a solid/void design [41]. The sensitivities are calculated analytically using the adjoint method [45].

The resulting designs (TOD's) from the optimization process, called 'as-designed', have a voxelized surface, because of the voxel-based discretization. Subsequently, the 'as-built' design was achieved by applying the 'Iso volume' filter to the intermediate design and exporting to an .stl file using vtk 9.1.0 [46]. The 'as-built' design was used for additive manufacturing and the subsidence risk assessment. Morphological properties, i.e. porosity, pore interconnectivity (isolated pores) and pore diameter distributions of the 3D implant designs, were quantified using the open-source python packages openPNM 3.1.0 [47] and porespy 2.2.2 [48].

### 2.4 Subsidence risk assessment

The subsidence risk assessment was carried out on the 'as built' geometry using a commercial explicit FE solver (Abaqus 2021, Dassault Systèmes, Vélizy-Villacoublay, France). Non-linear geometric effects and a general contact formulation were included. The general contact was modelled between the vertebrae and the implant with a tangential friction coefficient of 0.8 [49]. The bone tissue was modelled as a ductile material, using a rate-independent elasto-plastic material model that captures the asymmetric tension-compression yield and post-yield behaviour of bone [39] (Figure 4.b). Each FE model was meshed with ten-node tetrahedral elements (C3D10M), with an average element edge length of 2.0 mm for the vertebrae (selected after a mesh sensitivity study) and 0.5 mm for the cages. For both patients, three FE models were built to model the OTS, ACD and TOD cages, respectively. The models include the patient's FSU with the patient-specific calibration, segmentation, and material mapping (section 2.1, 2.2). A Follower Load (FL) of 310 N and 1250 N for patient 1 and 2 were selected from the study of *Rohlmann et al.* and the Orthoload dataset, based on body weight and representative of hyper-physiological loading for the patient [35], [50]. The FLs were applied uniformly distributed on the superior endplate of the superior vertebra, such that the direction of the load was aligned with the physiological alignment of the FSU (Figure 4.a) [51]. The inferior endplate of the inferior vertebra was constrained in all translations (Figure 4.a). The models were solved on a computational cluster using 64 cores in parallel.

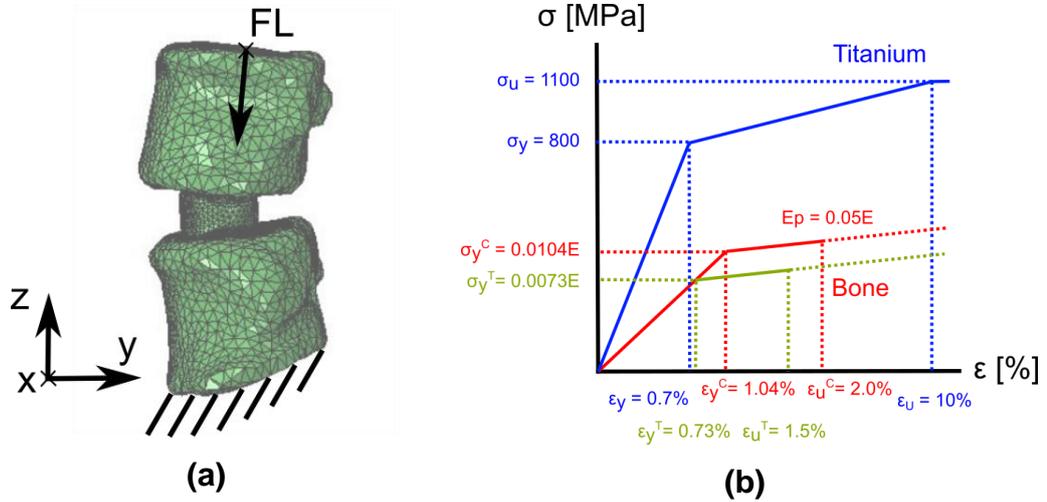

**(a)**                         **(b)**

*Figure 4, FE model for subsidence risk assessment a) FL application and support constraint are illustrated including the right-handed orthogonal coordinate system (data from patient 2) b) For the bone elements a tension yield strain $\varepsilon_y^T$ of 0.73% and a compressive yield strain $\varepsilon_y^C$ of 1.04% are assumed [39]. After reaching the yield point, a post-yield modulus $E_p$ of 5% of element's elastic Young's modulus E is implemented according to [39]. Subsidence risk was quantified using a maximum principal strain limit of 1.5% and a minimum principal strain limit of -2.0% [52]. For titanium, the yield and ultimate strain are assumed to be 0.7% and 10% [53] (figure is not to scale).*

For each FE model, subsidence risk was quantified as the degree of overloading of the vertebrae using a maximum principal strain limit of 1.5% and a minimum principal strain limit of -2.0% [52]. For each element '$i$' in the bone domain, the factor of fracture risk ($FFR_i$) [52] was calculated using:

$$FFR_i = max\left(\frac{\varepsilon_{max,i}}{1.50\%}, \frac{\varepsilon_{min,i}}{-2.00\%}\right)$$

where $\varepsilon_{max,i}$ is the maximum principal strain of element $i$, in the centroid of the element and $\varepsilon_{min,i}$ the minimum principal strain of element $i$, in the centroid of the element.

### 2.5   Mechanical testing and FE model validation

A prototype TOD design for patient 1 was additively manufactured on an EOS M100 (EOS GmbH, Robert-Stirling-Ring 1, D-82152 Krailling, Germany) using titanium powder with a particle size range from 10 to 45 μm (Osprey Ti-6Al-4V-ELI Grade 23 from Sandvik AB, SE-811, 81 Sandviken, Sweden). Fixtures were created to fit the TOD top and bottom surfaces (Figure 7.b). Mechanical testing was performed (n=6) using a static axial compression test (similar as described in ASTM F2077) on a Zwick Z100 (Figure 7.c). The FE model that is described in section 2.4 was used (Figure 7.d).

## 3   Results

### 3.1   Optimized cage

The resulting TODs for both patients are illustrated in Figure 5. A 'box-like' design emerges with porous walls and hollow in the inside but with several internal struts. The outside shape of both TODs differs due to the varying endplate shapes of the vertebrae. The top and bottom of the implants, parts that are in contact with endplates, are almost without pores. Furthermore, the internal structure and the porosity in the walls is significantly different for both TODs, meaning that both TODs have similar macroscopic properties i.e., maximum contact area and box-like design to satisfy the same goal (reducing subsidence risk) but achieve this with different patient-specific micro-structures. The discreteness measure reached close to the desired 3% value with 3.4% and 3.3% for patient 1 and 2, respectively. The designs satisfy all the constraints (see Appendix A for detailed optimization history plots). Morphological details of the TODs for both patients are listed in Table 4. The minimum strut size is achieved, and no isolated pores are found, which make the TODs manufacturable.

**Patient 1**  **Patient 2**

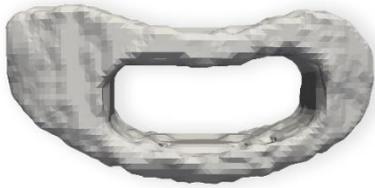

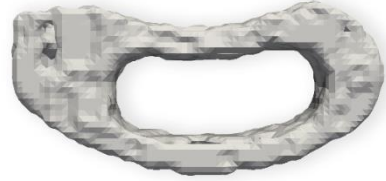

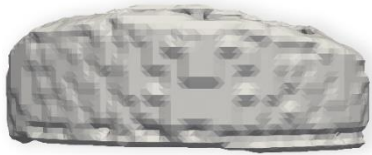

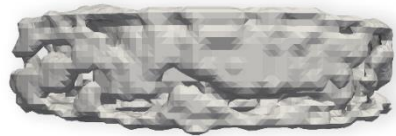

*Top and Side view*

*Top and Side view*

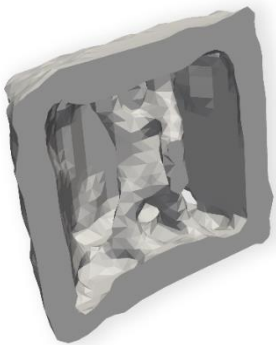

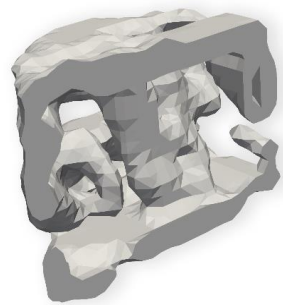

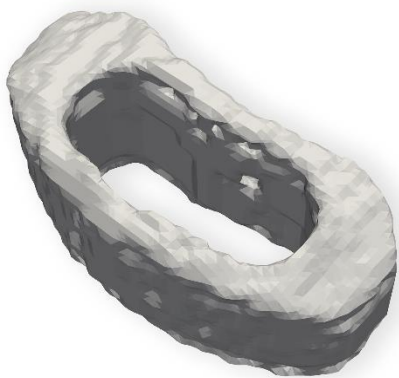

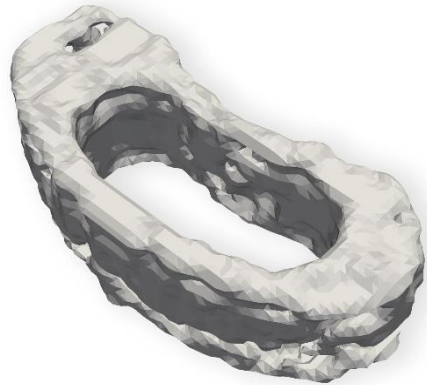

*Overview and cross-section, strut-structure is visible in large hollow pocket*

*Overview and cross-section, strut-structure is visible in large hollow pocket*

*Figure 5, TOD designs per patient (patient 1 on left, patient 2 on right)*

|  | **Patient 1** | **Patient 2** |
|---|---|---|
| **Porosity** | 59.3% | 51.1% |
| **Isolated pores**<br>**(a pore enclosed by solid material)** | 0 | 0 |
| **Mean pore size** | 4.7 ± 1.1 mm | 5.7 ± 1.4 mm |

## 3.2 Subsidence risk comparison

The maximum FFRs for each design are listed in Table 5. Histograms of the FFR values for the different designs are illustrated in Figure 6. We observe a decrease in subsidence risk from the OTS to the ACD and the TOD for both patients. Furthermore, the reduction in subsidence risk is higher for patient 2, the patient with normal bone quality. Specifically, for patient 1, subsidence risk is reduced by 54% compared to the OTS implant and by 91% for patient 2.

*Table 5, maximum FFR (factor of fracture risk) for three implants in patient 1 and 2: OTS (Off-the-shelf), ACD (Anatomically Conforming Device) and the TOD (Topologically optimized device).*

| **Patient ID** | **Patient classification** | **FFR OTS** | **FFR ACD** | **FFR TOD** |
|---|---|---|---|---|
| Patient 1 | Osteopenic | 3.7 | 2.2 | 1.7 |
| Patient 2 | Normal | 2.2 | 0.7 | 0.2 |

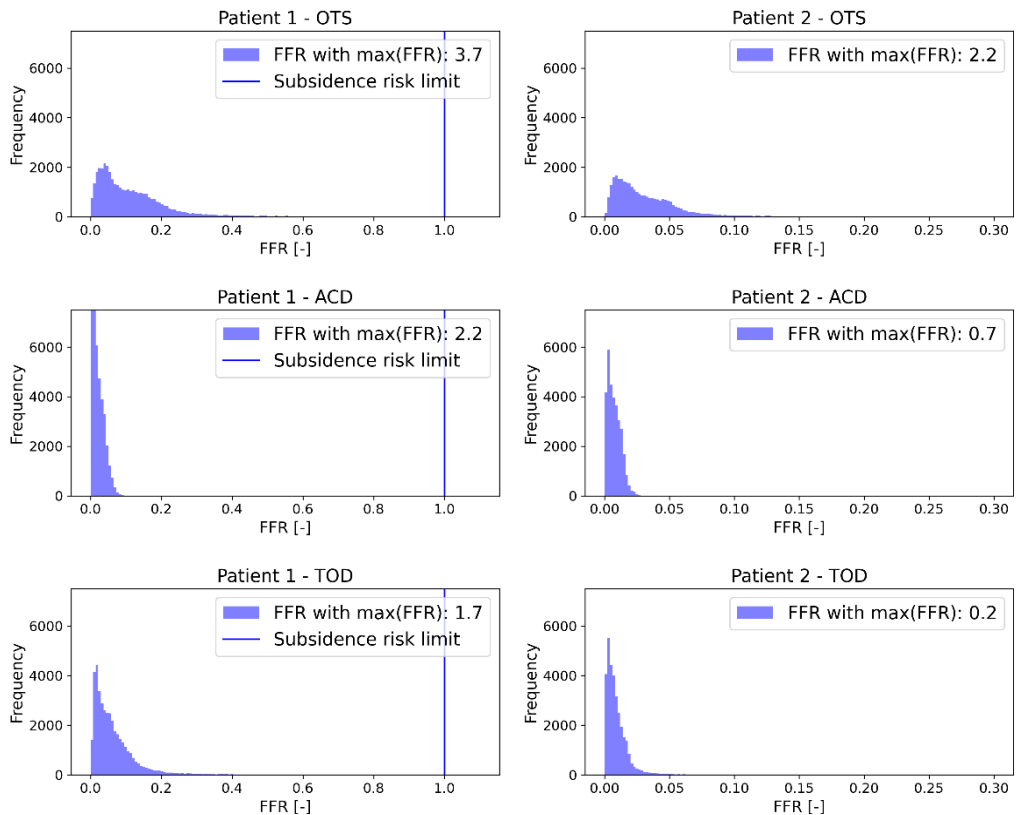

*Figure 6, histograms for patient 1 (osteopenic) and patient 2 (normal bone quality) for the OTS, ACD and TOD*

### 3.3 Mechanical tests and FE validation

Prototypes of TOD designs were successfully additively manufactured and the resulting cages (TODs) are illustrated in Figure 7.a. The force-displacement response from the mechanical testing and the FE simulations are illustrated in Figure 7.e and f. Due to a malfunction of the mechanical testing setup, one sample was excluded from the results. The TODs were loaded until failure, according to ASTM F2077, to test the integrity of the implant. We note that the physiological loading range, up to 1500N, is relevant for the validation of the subsidence risk FE model (Figure 7.f) and that the response in this range is slightly non-linear for both the FE model and the mechanical test results. Moreover, we observe a close match between the FE model and the mechanical test results.

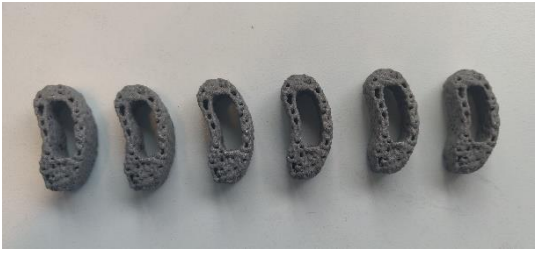

**(a)**

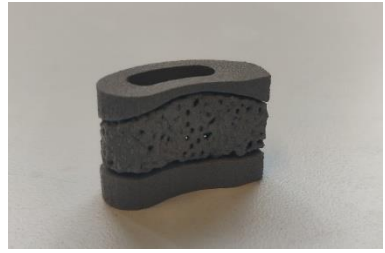

**(b)**

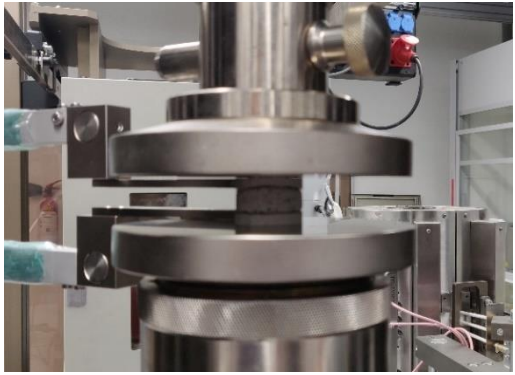

**(c)**

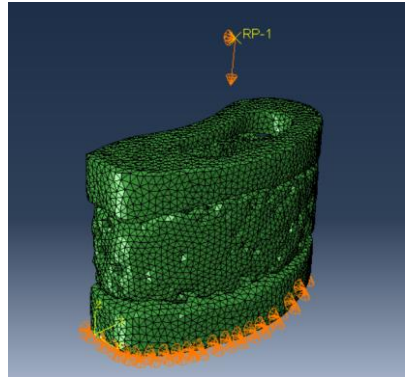

**(d)**

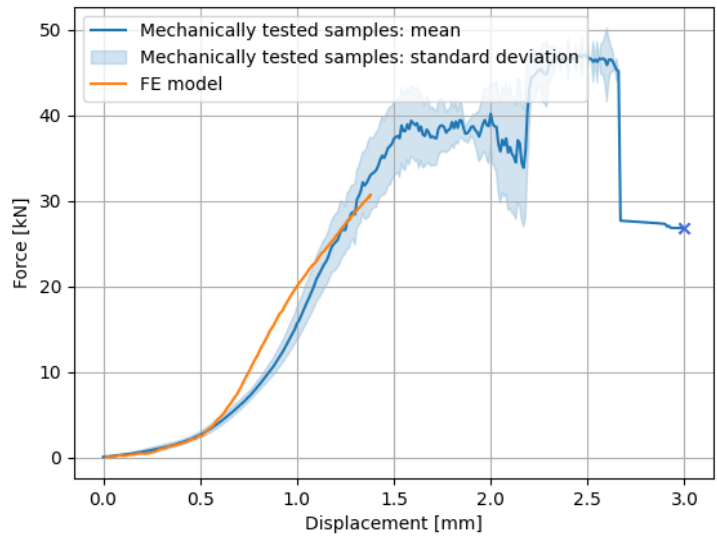

**(e)**

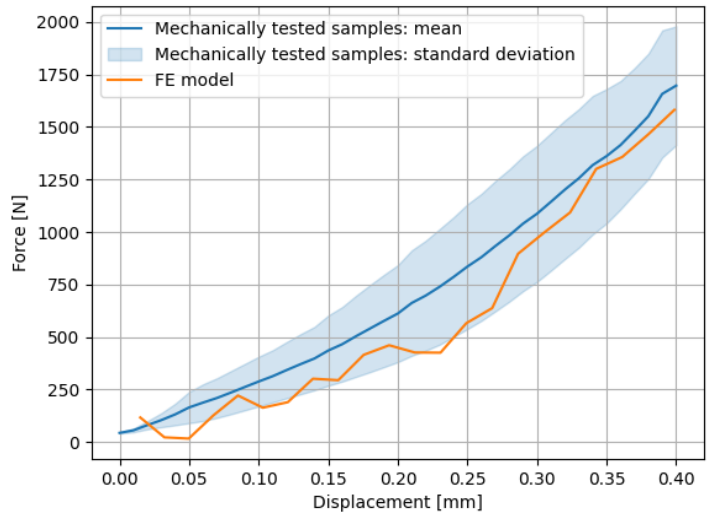

**(f)**

*Figure 7, prototyping, mechanical testing and FE validation a) TOD additively manufactured samples (data from patient 1) b) fixtures to facilitate the mechanical testing are demonstrated to fit the TOD c) the mechanical testing compression setup d) the FE model to mimic the mechanical testing e) Force-Displacement curve comparing the FE model and mechanical testing results. The FE model results are plotted till ultimate strain (10%) f) The physiological relevant part of the Force-Displacement curve is plotted up to 1500N.*

## 4 Discussion

The aim of this study was to optimize patient-specific spinal fusion cages using a novel full-scale TO formulation that is taking the structural response of the adjacent vertebrae into account in the optimization process. We hypothesized that the TODs would reduce subsidence risk compared to ACDs and OTS cages. We demonstrated that it is feasible to structurally optimize patient-specific spinal fusion cages resulting in anatomically and mechanically conforming devices. We found that subsidence risk (maximum FFR) was reduced by 91% for the TOD compared to the OTS for patient 2 (patient with normal bone quality). For patient 1 (patient classified as osteopenic) the subsidence risk was reduced by 54% with the TOD.

Comparing the subsidence risk with the ACDs and the TODs for patient 1, we observed a decrease from 2.2 to 1.7 (22.7%). For patient 2 the TOD could additionally reduce subsidence risk compared to the ACD from 0.7 to 0.2 (71.4%). This suggests that the effectiveness of the method may depend on patient-specific factors such as bone quality. Several authors investigated the correlation between the effectiveness of a fusion cage and patient-specific factors such are bone quality. Lund *et al.* showed the increased stabilising effect of OTS cages with increasing vertebral bone density [37]. Jost *et al.* reported the increasing compressive failure load of vertebrae, tested with several OTS cages, with increasing vertebral BMD [54]. Further research needs to be carried out to verify this, however, we do observe that the load-bearing capacity of the vertebrae adjacent to the implant for the osteopenic patient (patient 1) is exhausted even after the optimization. This decreases the design freedom of the optimizer in terms of finding an implant design that reduces strain concentrations. To explore this speculation, we added more design freedom to the optimizer by increasing the design domain for patient 1. The maximum FFR values did indeed reduce compared to the implant with the design domain that is limited to the OTS implant geometry. This suggests that allowing the implant sized to vary (patient-specific implant size), could be beneficial and increase the effectiveness of the TO process, as intuitively can be expected.

We found our hypothesis on reducing subsidence risk, by reducing strain concentrations in the adjacent vertebra, to hold. However, we also found that maintaining the highest possible contact area between the implant and the adjacent vertebra is important. The working principal of reducing subsidence risk for the ACD is clearly to increase the contact area of the implant to the vertebrae endplates. This is aligned with the observations of previous studies [8], [9]. The unique benefit of using TO is that porosity is introduced, so that subsidence risk can be reduced, without compromising the primary stability of the implant.

Two manufacturability constraints were incorporated in the TO process, the minimum strut diameter control (robust formulation) and the connectivity constraint. Adding these manufacturability constraints resulted in manufacturable designs. Additionally, the implant's integrity was confirmed by comparing the implant's compression yield limit, which was found to be higher than 20000 N to the yield limit data gathered by the FDA from 79 TLIF devices with a compression yield limit of 17479 ± 4301 N [55].

The mechanical test results and FE validation (Figure 7.f) show that the FE model could reproduce the mechanical tests. We conclude that the mechanical tests could validate the FE model (excluding the bone material model, which was validated in [39]) that was described in section 2.4.

Several assumptions were used in this study which are worth mentioning. First, in case of the TO formulation, a fully bonded bone-implant interface (perfect bonding of the design domain with the two vertebrae bodies) and a linear elastic FE and linear stress-strain relationship were assumed. This assumption is valid since daily-living loading is applied that cause strains in the linear regime. However, for the subsidence risk assessment, a contact formulation, non-linear FE, and a non-linear material model were used, which are needed to capture plasticity effects to model subsidence risk. Second, it is assumed that errors due to the clinical CT scan resolution and material mapping will not influence the results. Third, the authors assume that the factor of fracture risk (FFR) is a proper representation of actual subsidence risk and that using principal strains is a better field variable than, for example, von Mises stresses for applications including biomechanical material mappings. Using the principal strains allows for introducing asymmetric tensile–compressive strain constraints in the TO process, the FE model and in calculating the FFR.

A limitation of using high-resolution TO in the design process is the large computational cost. The TO process was run in parallel on 1000 cores and took 24 hours to converge to an acceptable solid/void design (discreteness measure below 3%). The authors argue that a run time of 24 hours qualifies the method as clinically feasible. On the other hand, restricting the run time to 24 hours limits the mesh resolution of the TO domain. Further research needs to explore ways to increase resolution and reduce the computational cost *e.g.,* trough model reduction methods, techniques using physically informed neural networks [56] or use TO on GPUs [57]. Increasing the mesh resolution will increase the design freedom of the optimizer. This will potentially reduce the subsidence risk caused by the TOD. A second limitation is that the current mesh size to discretize the vertebrae in the subsidence risk evaluation is potentially too coarse to fully capture the Young's modulus gradient between the trabecular bone and the vertebra endplates. In this study, we accept this inaccuracy because all the vertebrae are modelled the same way. Finally, a limitation is associated with the clinical implementation of the method. An unsolved clinical problem is the precise free-hand positioning of the implant during surgery. Since the TOD is optimized for a specific position compared to the adjacent bone structures, imperfect placing of the TOD would possibly reduce the effect of the TOD in terms of lowering subsidence risk. Computer navigation and surgical instruments that facilitate precise positioning could mitigate this.

## 5 Conclusion

This study demonstrates a design and optimization method, using a novel full-scale topology optimization formulation that can optimize patient-specific spinal fusion cages, taking the mechanical response of the adjacent bone structures into account. This results in anatomically and mechanically conforming devices. The presented workflow produced manufacturable implant designs that can be additively manufactured in medical grade titanium. Real patient data was incorporated. The method is effective at lowering subsidence risk for a patient classified as osteopenic (54% reduction compared to an off-the-shelf implant) and a patient with normal bone quality (91% reduction compared to an off-the-shelf implant). The method's effectiveness depends on patient-specific factors, with the implant size being identified as one of the influencing factors. The authors presume that the novel patient-specific topology optimization approach, upon further development, can become an essential tool in the implant design process to reduce risks of implant induced endplate fracture, reducing reoperation rates, improving patient care and ultimately, reduce the total cost burden of the health-care system.

**Acknowledgements**

We would like to thank Dave O'Riordan (Schulthess Klinik, Zürich) for his efforts to provide the CT data.

**Disclosure statement**

The authors declare that they have filed a patent application based on the topology optimization method that is used in this work.

**Data availability**

**Funding**

This project has received funding from the European Union's Horizon 2020 research and innovation programme under the Marie Sklodowska-Curie Grant Agreement No. 812765.

**Appendix A**

**Optimization history plots**

The optimization history highlights the stability and robustness of the optimization process (Figure A1). The initial homogeneous design converges to the final design with 'steps' caused by the increase of β (Figure A1.a and b). The increase of the objective function towards the final design indicates that the bone-implant system is reducing stiffness to satisfy the constraints (Figure A1.c, d, e, f, g, h). The compression load-case (lc. 3) is dominant for the maximum and the minimum principal strain constraints (Figure A1. e, f, g, h). The discreteness measure is converging close to the desired 3% value (Figure A1.i and j).

**Patient 1**                    **Patient 2**

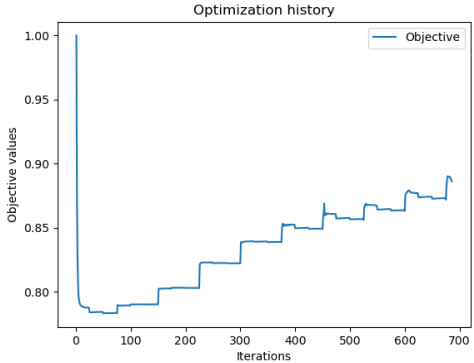

**(a)**

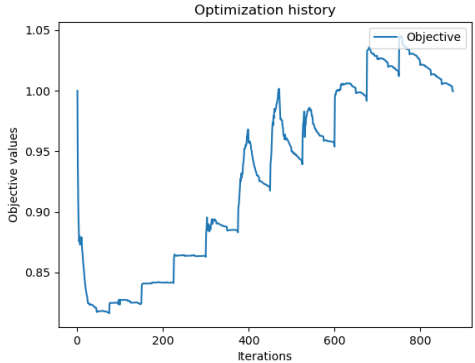

**(b)**

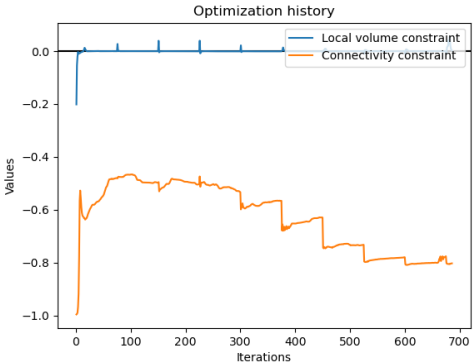

**(c)**

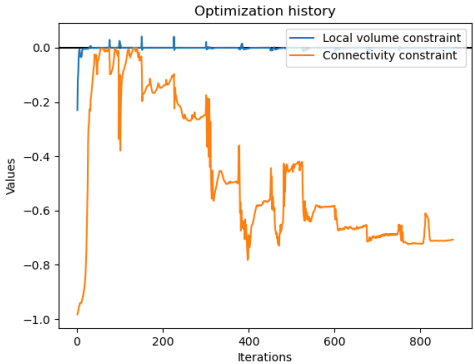

**(d)**

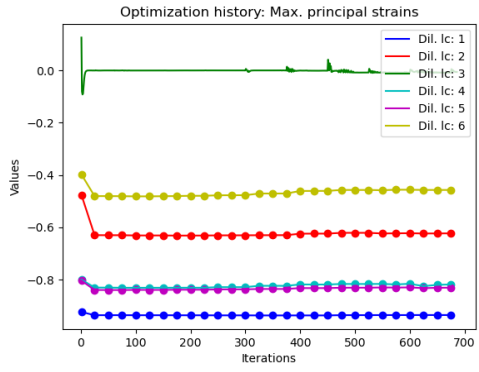

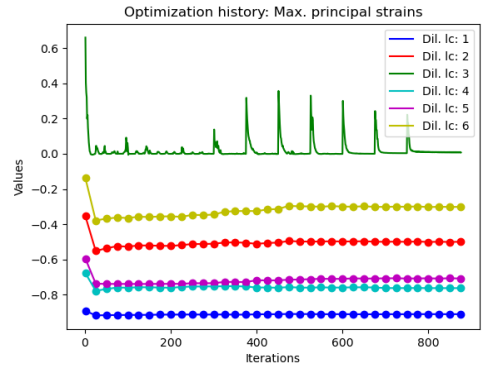

**(e)**
**(f)**

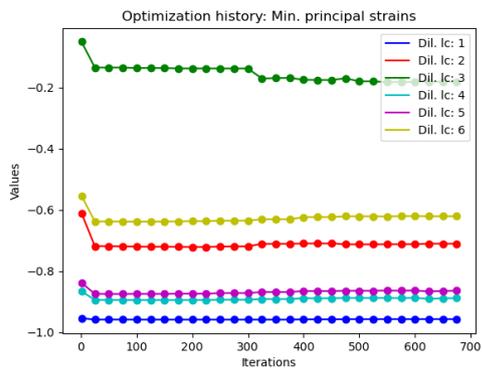

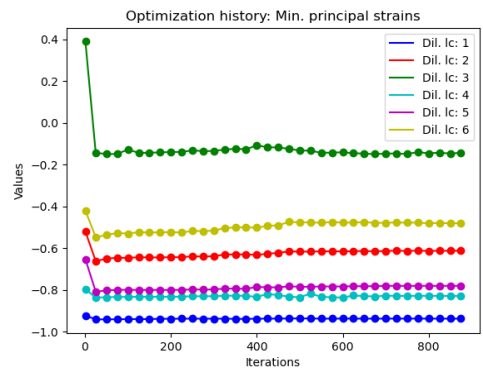

**(g)**
**(h)**

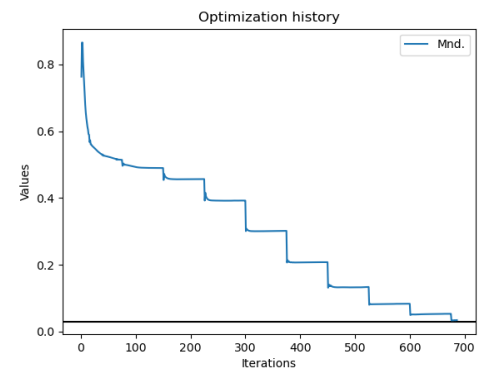

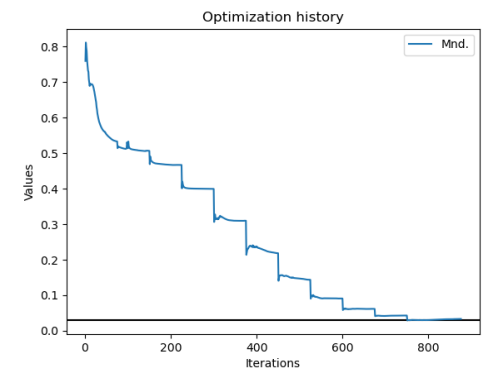

**(i)**
**(j)**

*Figure A1, optimization history plots for patient 1 and 2*